\documentclass[preprint]{aastex}
\usepackage{natbib}
\slugcomment{Accepted by Astrophysical Journal, August 2, 2000}

\begin{document}

\shortauthors{Lommen et al.}
\shorttitle{New Pulsars from an Arecibo Drift Scan Search}

\newcommand{\Msolar}{{$\rm\ M_\odot$}}
\newcommand{\Mearth}{{$\rm\ M_\oplus$}}
\newcommand{\DEG}{^{\circ}}
\newcommand{\de}{$^{\circ\ }$}
\newcommand{\nde}{$^{\circ}$}
\newcommand{\BAlph}{{$\alpha\ $}}
\newcommand{\nAlph}{$\alpha$}
\newcommand{\pr}{$^\prime\ $}
\newcommand{\dpr}{$^{\prime\prime}\ $}
\newcommand{\bet}{{$\beta\ $}}
\newcommand{\nbet}{{$\beta$}}
\newcommand{\hz}{{\rm\ Hz}}
\newcommand{\Mhz}{{\rm\ MHz }}
\newcommand{\tria}{$\triangle$}
\newcommand{\diam}{\Large $\diamond$}
\newcommand{\exs}{{\large $\times$}}
\newcommand{\hou}{$^h$}
\newcommand{\minu}{$^m$}
\newcommand{\seco}{$_{\cdot}^{s}$}
\newcommand\etal{{et al.~}}
\newcommand\eg{{\it e.g.,~}}
\newcommand\Msun{M$_\odot$}

\title{ NEW PULSARS FROM AN ARECIBO DRIFT SCAN SEARCH}

\author{ Andrea N. Lommen\altaffilmark{1}, 
         Alex Zepka\altaffilmark{1}\altaffilmark{2}, 
         Donald C. Backer\altaffilmark{1},
         Maura McLaughlin\altaffilmark{3},
         James M. Cordes\altaffilmark{3},
         Zaven Arzoumanian\altaffilmark{4},
         Kiriaki Xilouris\altaffilmark{5}}

\altaffiltext{1}{Astronomy Department \& Radio Astronomy Laboratory,
University of California, Berkeley, CA 94720, email: alommen@astro.berkeley.edu, dbacker@astro.berkeley.edu}
\altaffiltext{2}{Currently at Cadabra Design Technology,
3031 Tisch Way, Suite 200, San Jose, CA 95128, email: zepka@cadabradesign.com}
\altaffiltext{3}{Department of Astronomy, Space Sciences Building, Cornell University, Ithaca, NY 14853; mclaughl@spacenet.tn.cornell.edu, cordes@spacenet.tn.cornell.edu}
\altaffiltext{4}{Center for Radiophysics and Space Research, 
Cornell University, 
currently at NASA-Goddard Space Flight Center,
Mailstop 662.0,
Greenbelt, MD 20771, email: zaven@milkyway.gsfc.nasa.gov}
\altaffiltext{5}{University of Virginia, Astronomy Department,
P.O. Box 3818, Charlottesville, VA 22903, email:  kx8u@carina.astro.virginia.edu}

\begin{abstract}

We report the discovery of pulsars J0030+0451, J0711+0931, and
J1313+0931 that were found in a 
search of 470 square degrees at 430 MHz
using the 305m Arecibo telescope.  
The search has an estimated sensitivity 
for long period, low dispersion measure, low zenith angle, and
high Galactic latitude pulsars
of $\sim$ 1 mJy, 
comparable to previous Arecibo searches.
Spin and astrometric parameters
for the three pulsars
are presented along with polarimetry at 430 MHz.
PSR J0030+0451, a nearby pulsar with a period of 4.8 ms,
belongs to the less common category of isolated millisecond pulsars. 
We have measured significant polarization in
PSR J0030+0451 over more than 50\% of the period,
and use these data for a detailed discussion of its magnetospheric geometry.
Scintillation observations of PSR J0030+0451 provide an estimate of the 
plasma turbulence level along the line of sight through the local 
interstellar medium.
\end{abstract}


\section {Introduction}

\setcounter{footnote}{0}

Pulsar searches work toward completion
of the Galactic inventory of active neutron stars, and
provide new tests of both the validity and robustness of
emission models and ideas concerning the origin of pulsars.
During the time of the Gregorian Dome upgrade at the
Arecibo Observatory, 
a joint effort by several collaborations was aimed at
surveying the sky visible from Arecibo 
(declination range roughly $-$1\de to 39\de) 
in search of new pulsars.   
A total of 44 pulsars including 5 MSPs have been discovered so far
by the other institutions \citep{Foster95, Camilo96, Ray96}.
The Berkeley/Cornell effort was divided into two independent
areas of the sky, in which a total of 5
more pulsars have been discovered: 
2 are reported by \citet{McLaughlin99}, and 3 are reported here.

This paper presents the discovery
of the 9th isolated millisecond
pulsar (MSP\footnote{By convention `MSP' refers to an object having
both a millisecond period (P $<$ 100 ms) and a low period
derivative ($\dot{P} \lesssim 10^{-19}$ s s$^{-1}$).
Young pulsars
with periods of tens of milliseconds are therefore excluded.}) in the
disk of the Galaxy 
and two additional slow pulsars.
We investigate the new pulsars in terms of
the Rotating Vector 
Model (RVM) \citep{Rad69} and models of core vs. cone 
emission \citep {Rankin83}.
The new isolated MSP adds to the 8 other known objects in this category
that require disruption of their companions if they achieved millisecond
periods through accretion of matter and angular momentum from a companion star
\citep{Bhattacharya91, Phinney94, vandenheuvel95}.

We present the details of the search in \S 2, the confirmation
process in \S 3, and our observations
 and the resulting new pulsar timing models in \S 4.
In \S 5 we present both intensity and polarization profiles
of the new objects
and use the results for a detailed 
study of the magnetospheric geometry of PSR J0030+0451.
In \S 6 we summarize the results and briefly discuss MSP evolution. 

\section {Arecibo Drift Search for Pulsars}

For the purposes of the joint search effort, 
the sky was divided into declination
strips of 1$^{\circ}$. The Berkeley/Cornell team was assigned the
declinations centered at $-0.5^{\circ}$, $4.5^{\circ}$, $9.5^{\circ}$,
$14.5^{\circ}$, $19.5^{\circ}$, $24.5^{\circ}$, $29.5^{\circ}$ and
$34.5^{\circ}$. 
We are reporting here on
a total of 470 square degrees of these strips that
were successfully surveyed with
a manageable level of
radio frequency interference (RFI) between 1994 October and 1995 February.

During the upgrade construction the azimuth arm was fixed, but
observations were still possible as the sky drifted past the
parked receiver.
We used the 430-MHz line feed receiver with 8 MHz bandwidth which was
divided by analog filters into 32 channels. 
The detected powers from orthogonally polarized channels were summed and
Nyquist sampled at a 4-kHz rate using the ADAGIO data acquisition system.
The observations consisted of long integrations taken with the
telescope at a fixed azimuth and elevation.  Because a pulsar transits the
telescope beam ($10^{\prime}$
half-power width) in about 30 seconds, we analyzed the data in blocks 
of $2^{17}$ time
samples ($\sim 32$s) with each block sensitive to roughly a 
$8^\prime \times 10^\prime$ 
region of the sky.

In the first column of Table 1 we list in square degrees the total sky area
that was relatively
RFI-free for each declination strip.
We obtained no data for strips centered at $14.5^\circ$ and
$19.5^\circ$. About 41\% of the area covered was observed more than
once.  
The area observed in each declination strip 
is given in the second column
of Table 1.  In parentheses are the percentages corresponding to the fraction
of the strip covered in our survey, and the area observed more
than once is given in column 3. 
There was an abundance of data taken at  9\de declination
as a result of the telescope being essentially fixed there during the
final days before telescope operations were completely halted for the upgrade. 

We processed 
the data using essentially the same well-tested
software from previous pulsar searches performed with the same
instrumentation \citep[]{Zepka96}.
Our analysis used clusters of workstations at Berkeley and 
Cornell.  We discarded pulsar
candidates from observations in which strong 
RFI was detected.
The discarded fraction was nearly $30\%$ of the total.

The off-line data processing, which is described more fully in 
\citet{Zepka96},
consists of removing trial dispersive delays followed by fast Fourier
transforming and harmonic summing to look for maxima in frequency.
The harmonic summing (up to 8 harmonics per candidate) increases
the sensitivity of the search more for long period pulsars, so the
sensitivity limit to the shortest period MSPs will be 
higher than the 1 mJy we have quoted.  See \nocite{Ray95, CNT96, Foster95} 
Foster et al. 1995, Ray et al. 1995, and Camilo, Nice, \& Taylor 1996
for curves showing typical sensitivity vs. pulse period.

Three candidate pulsars were confirmed in the analysis
of the drift scan data.
All three were detected at the same Dispersion Measure (DM) and position on
at least 3 days and with a similar profile on each of the days.
In addition to these 3 pulsars
we selected the top 22 candidates detected by the search code
which had a significance at least 6$\sigma$ above the noise and which
were also promising candidates based on either the cleanliness of
the data (amount of RFI present) or their detection on multiple days.
These 25 candidates were reobserved in targeted searches 
during the confirmation run described in \S 3.  

A rough estimate of the sensitivity of the search is obtained by
using the following expression
which gives the minimum detectable 
flux at 10$\sigma$ 
for a pulsar at high Galactic latitude and at the beam center for zero zenith
angle:
$$
S_{{\rm min}}= {\frac{10T_{{\rm sys}}}{G\sqrt{n\Delta\nu\tau}}}\sqrt{\frac{W}{P-W}}= 0.6\ \rm{mJy}, \eqno(1)
$$
where $G = 17 $ K Jy$^{-1}$ is the effective gain of the Arecibo telescope
at $430\Mhz$,
$n = 2$ is the number of polarizations 
summed, $\Delta\nu = 8\Mhz$ is the total observing bandwidth, 
$T_{\rm sys}=70\ K$ is the
system temperature, $\tau = 32\ $s is the integration time, 
$P$ is the period of
the pulsar and $W=0.1P$ is a typical pulsar duty cycle, including
effects of dispersion smearing within the channel bandwidth and scattering.  
The minimum flux increases monotonically with zenith angle, which means
that for declinations away from $\sim 18 \DEG$ (the latitude of the Arecibo
Observatory) the search is less sensitive 
(see Camilo, Nice, \& Taylor 1996 \nocite{CNT96} for
gain vs. zenith angle at the Arecibo Observatory using the 430-MHz linefeed).  
The maximum detectable period
of the search was set by software to 5s.
We estimate the overall sensitivity of the search to
be approximately 1 mJy.
There is a large uncertainty in the completeness of the search 
at the 1 mJy limit owing
to the ``clean" candidate selection process, the favoritism
given to multiply observed candidates and to high numbers
of harmonics, and the systematic
effects related to zenith angle.  

In addition, we detected
the previously known pulsars J0621+1002 (28ms), B0626+24 (0.48s),
J1022+1001 (16ms), B1541+09 (0.75s), B1633+24 (0.49s), B1726$-$00
(0.38s) and J2222+2923 (0.28s), the first 6 of which have flux 
densities of
7, 27, 14, 54, 9, and 10 mJy respectively, 
while the flux of J2222+2923 is unknown. 
The following pulsars were missed, i.e. their positions are in a region
that was searched at least once and we did not detect them:
J0051+04 (0.36s), J0533+04 (0.96s), J1834$-$0010 (0.52s),  J1913+0935 (1.2s),
and J1926+0431 (1.1s).
The fluxes of the last three are 5, 22, and 3 mJy, respectively, while
those of the first two are unknown.  The new pulsars have flux densities of 
2 to 8 mJy as reported below.

One measure of the sensitivity of a search is the total number
of pulsars, new and previously known, detected per area.
Our search has similar sensitivity to the
3 other declination strip surveys \citep{Ray95, Foster95, Camilo96},
which had a detection rate 
ranging from 1 pulsar per 36 to 46 sq. deg., with an average
detection rate of 1 pulsar per 41 sq. deg.
We detected 1 pulsar per 47 sq. deg.

\section {Confirmation of Pulsar Candidates }

In 1997 December the Arecibo Observatory was just seeing first light after the
upgrade.  
Our initial observation made use of the 430-MHz line-feed receiver on 
Carriage House 1, the same
system used in the drift search.
The entire platform was higher than normal which
led to a pointing uncertainty of about 0.5$^\prime$
and a shift in the center frequency of the 430-MHz line-feed
to 433 MHz.  
With the telescope in this state we made our confirmation
run of the 25 candidates from our search.  
For each of the 25 candidates we took a 30-minute time-series data set 
using the Penn State Pulsar Machine which has 128 channels 
covering 8 MHz of bandwidth and uses a sample time
of 80\ $\mu$s \citep{Foster95}.

Of the 25 candidate pulsars gleaned from our search, only
the ones detected on multiple days were confirmed:
J0030+0451 (4.8ms), J0711+0931
(2.4s), and J1313+0933 (0.85s).  
PSR J0030+0451 was also detected independently by 
D'Amico (2000)\nocite{Damico00}. 
A fourth candidate which had been seen on multiple days of the search
(J0814+09, 0.65s) was 
not confirmed in the 1997 December observations discussed, and has
not been confirmed thus far.
We looked for each
of the remaining candidates on at least one other night, and in most cases
two nights, without detecting any additional pulsars.  

\section{Timing Models for the New Pulsars}

Follow up observations of the 3 new pulsars
were conducted over a period of 2.3 years from 
1997 December through 2000 March at the
Arecibo Observatory using
the Arecibo-Berkeley Pulsar Processor.  This 
signal processor provides coherent dispersion removal
within each of 32 channels; see \citet{Backer97} for a partial technical 
description. The dispersion removal is done
via complex convolution of voltage samples in the time domain, 
using up to 1024 time samples for
each channel for channel bandwidths less than or equal to 1 MHz.
For intensity data one can
use up to 3.6-MHz bandwidth per channel, 
making available 112-MHz overall bandwidth.
Full Stokes parameters can be
acquired for channel bandwidths up to 0.875 MHz. 

The timing data were cross correlated with a template, 
and the resulting TOAs analyzed using the TEMPO software 
package\footnote{http://pulsar.princeton.edu/tempo}.
Table 2 provides spin, astrometric and other parameters for each of the 
3 new pulsars.  
Finding the timing solution was challenging.  Though we have
more than 2 years of data on each pulsar, the time series are sparsely sampled.
For PSR J0711+0931 we have 7 independent epochs of data, 
for J1313+0931
we have 5, and for PSR J0030+0451 we have 9, an epoch being defined by data
separated from other data by two weeks or more.  
However, there are several epochs for each object, in which
we have three or more days of data spread over a week.   
These epochs were critical for finding
an overall timing solution because we obtain a phase connected 
solution within the epoch, i.e. a
period, $P$,  and apparent period derivative for that epoch, 
$\dot{P}^\prime$. 
$\dot{P}^\prime$ differs
from the intrinsic period derivative 
owing primarily to $\sim1^\prime$ position errors.
Connecting the 
$P$ and $\dot{P}^\prime$ from different epochs then required
finding this position offset, a simple task of 
fitting a sine wave
to $P$ vs. epoch, whose derivative must be $\dot{P}^\prime$ vs. epoch.
Having found the position in this manner to within $\sim 1^{\prime\prime}$,
we were then able to phase connect the entire data set
and fit for the timing parameters shown in Table 2.

The root mean square (RMS) of the timing residuals quoted in Table 2
are from fits to the combined 430-MHz and 1.4-GHz data sets.   
The uncertainties in Right Ascension, Declination, $P$, and $\dot{P}$, are
estimated by 
doubling the errors given by TEMPO, an ad hoc
procedure that attempts to  account for timing noise, if any, and 
other nonrandom sources of error in the fits.

The $P$ and $\dot{P}$ for PSR J0030+0451,
4.8 ms and 1 $ \times 10^{-20} $\ s s$^{-1}$ respectively,
are near the 
``death line"\footnote{See, e.g., Lorimer (1999)\nocite{Lorimer99} and
Zhang, Harding, \& Muslimov (2000)\nocite{zhang00}
for discussions on the pulsar death line.}  
of the MSP distribution (see \S\ref{sec:disc} for further discussion).
The $P$ and $ \dot{P} $ of
the two slow pulsars fall centrally in the distribution of slow
pulsars. 
PSR J0711+0931 is close to the plane of the Galaxy, while the other
two are out of the plane.  The 430-MHz luminosity of PSR J0030+0451 contends
with two other pulsars for the lowest luminosity of the known MSPs:
PSRs J1744-1134 and J2124-3358 at 0.4 $\pm$ 0.2 mJy kpc$^2$
\citep{Bailes97}. 

PSR J0030+0451 is the 4th closest known MSP, 
and is potentially one of the most observable MSPs as an X-ray source.
There is evidence that X-ray flux scales as 
$\frac{\dot{E}}{D^2}\propto\frac{\dot{P}}{P^3 D^2}$, 
where $\dot{E}$ is the energy loss rate, and $D$ is the distance to
the pulsar \citep{Becker97}.  By this relation PSR J0030+0451 is
expected to be the second brightest MSP in X-rays.
\citet{Becker00} report an X-ray detection using ROSAT.
PSR J0030+0451 is the 11th MSP detected in X-rays
and displays an X-ray profile similar to the 
radio profile.

We were able to fit for the proper motion of PSR J0030+0451
( $\mu_{\alpha} = -20 \pm 13 $ mas yr$^{-1}$ and 
$\mu_{\delta} = 32 \pm 31 $ mas yr$^{-1}$), but  
this has limited significance owing to the two remaining degrees
of freedom (9 epochs minus 7 free parameters including phase).
The fit is actually more meaningful for the upper
limit of 70 mas yr$^{-1}$ that it places on the proper motion.
Additionally, the measured $ \dot{P} $ places an upper limit
on the transverse velocity of the pulsar via the 
Shklovskii effect \citep{Shklovskii70, Camilo94}.  In
the case of PSR J0030+0451 the Shklovskii limit
is 60 mas\ yr$^{-1}$, similar to
the upper limit from the proper motion fit.  
A typical space velocity
for a recycled pulsar is 80 km s$^{-1}$
\citep{cordes97}, 
which corresponds to about
75 mas yr$^{-1}$ for PSR J0030+0451 at a distance of 230 pc.  
Therefore most of the pulsar's apparent $ \dot{P}$ may be due
to its proper motion. 

\section {Profiles:  Intensity and Polarization}
In addition to total intensity measurements of pulse profiles
at 430-MHz and 1.4-GHz obtained during timing programs,
polarization observations were made in 1998 June at the Arecibo
Observatory with the 430 MHz line-feed receiver on Carriage House 1.
See \citet{Sallmen98} for information on techniques employed.  
The data were not corrected for instrumental polarization, which
would at worst convert 10\% of linear power into circular power
\citep{Weisberg79, Stinebring84-1}.
Relative phase between the two circular polarizations
was measured by injecting a linearly polarized noise source into
the receiver after the feed.
The raw polarization data (orthogonal circular polarizations)
were converted to absolute flux units
using the standard bootstrap calibration method of determining the
flux of
a pulsed noise source by comparison with
an unpolarized standard continuum source, and then calibrating the pulsar flux
using the flux of the pulsed noise source. 
For standard flux calibrators at 430 MHz and 1.4 GHz we used nearby sources 
PKS 0829+18 (0.71 Jy at 410 MHz,
Wright \& Otrupcek 1990\nocite{Wright90})
and J0149+0555 (0.91 Jy at 1420 MHz, 
Condon {et~al.} 1998\nocite{Condon98}).

\subsection {PSR J0711+0531}\label{sec:0711} 
PSR J0711+0931 is a slow pulsar with P=1.2 s that is 
close to the Galactic plane 
($b=$8.8\nde)
and at a distance of $\sim$ 2.5 kpc which is
calculated from the DM using
the model of \citet{TaylorCordes93}.
At 430\Mhz this pulsar displays two components, while at 1.4 GHz there
is a single component (Fig. 1).  
The bifurcation of component structure as frequency is decreased
is indicative of ``conal" emission \citep{Rankin83}.  
However our data are not sufficient to determine whether the
peak actually bifurcates, or whether a new component becomes visible 
at lower frequencies.   Better frequency coverage
between 430\Mhz and 1.4 GHz 
is required to make this determination.
The DM quoted in Table 2 is 
calculated 
by lining up the single peak of the 1.4-GHz template profile
with the main peak of the 430-MHz template profile.
The small dip in intensity just before the peak is a real-time
signal processing error that appeared in our data around the time
these data were taken.   This dip is also visible in Figure 2 
(PSR J1313+0931),
and does not represent any phenomenon intrinsic to the pulsar.

The polarization position angle swing in PSR J0711+0931 is fit
well by the rotating vector model (RVM), but due to the limited 
extent of significant polarization in this profile,
there is no unique determination 
of \nAlph, the dipole magnetic field inclination angle with 
respect to the rotation 
axis, and \nbet, the separation of the line of sight with
respect to the magnetic axis at closest approach.  
We show a likely RVM in Figure 1 that results from
setting $\alpha = 60.0 \DEG$, the a priori most likely inclination
angle, and then finding the best-fit value of $\beta$ which is $ 7.2 
\DEG \pm 2.0 \DEG $.
The data are fit equally well with any $\alpha$ and $\beta$
with equivalent $\sin(\alpha)/\sin(\beta)$, the steepest
slope of the PPA swing.  The center of
symmetry of the fitted RVM falls 
between the two 430-MHz peaks, but mostly by virtue of that being the center of
the available polarization data.
This pulsar falls on the long-period, small spin-down end of the pulsar
population, which is where most of the conal-profile (single
and double) pulsars are found \citep{Rankin83}.

\subsection{PSR J1313+0931}\label{sec:1313}

PSR J1313+0931 is a single-peaked pulsar with P=0.85 s.  The
pulse peak is asymmetric, with a FWHM of approximately 6\de at both
1.4 GHz and 430 MHz .
The polarization data show a
clear position angle
sweep through the main pulse peak (Fig. 2) 
with a $ 90 \DEG $  discontinuous transition, 
indicative of an ``orthogonal mode" (suggested rotation shown with open
circles).
Setting \BAlph to $ 60.0 \DEG$,
$\beta $ fits to
$ -9.5\DEG \pm 3.0\DEG $ as shown in the figure, 
but the data are fit equally well with any $\alpha$ and $\beta$
where the slope $\sin(\alpha)/\sin(\beta)$ is the same.  

We note that in addition to the presence of extended emission on the
trailing side of the pulse, the center of the polarized emission falls 
toward the trailing edge of the main pulse.
These characteristics imply
that this profile in fact represents a double-peak with the 
second peak suppressed in intensity compared to the first.  
\citet{Lyne88} note a tendency for the
leading component to be the stronger of the two, and suggest that 
the emission process is somehow more effective at the leading edge
of the polar cap.  We see this effect in
PSR J0711+0931 and possibly PSR J0030+0451 as well 
(see \S\ref{sec:0711} and \S\ref{sec:0030polar}).

\subsection {Millisecond Pulsar J0030+0451}

The pulse profiles of PSR J0030+0451 at 430\Mhz and 1.4 GHz are
presented in Figure 3 to show the evolution of the pulse profile
with frequency.  
A pulsar displaying an interpulse like PSR J0030+0451 has 
two distinct possible geometries:
nearly aligned, with the interpulse resulting from the second crossing of
a wide-angle hollow ``cone" of radiation, or 
orthogonal with the two emission regions coming from opposite poles
\citep{Rankin83, Lyne88}.  The question of the basic geometry of
interpulse pulsars has been debated extensively \citep{Rankin83, Lyne88}, 
with most interpulse MSPs exhibiting qualities of both orthogonal and aligned
rotators (see, e.g. Xilouris et al. 1998\nocite{Xilouris98}).
Knowing which model was correct for each
pulsar would greatly enhance our knowledge of the geometry of
the underlying emission geometry.
The simplest test of geometry is that if
the pulse-interpulse separation changes with frequency we suspect we
are seeing a single cone of emission, with a frequency dependent 
emission height.  Unfortunately this test is not possible with PSR
J0030+0451 because both the pulse and interpulse are made of at least
two components 
whose positions in frequency cannot be identified using
our two-frequency data due to the complexity of the pulse shape.

In an attempt to quantify the evolution of the profile with frequency,
we fitted the 1.4-GHz average profile to a 6-Gaussian-component model
using the software of \citet{Kramer94I} and 
\citet{Kramer94II}.  
The Gaussian parameters of these components are given in Table 3.
The amplitude of component 2 has been set to 1.0 arbitrarily and
all other component amplitudes are measured relative to component 2.
The components chosen
are {\em not} uniquely determined by the data, but are chosen in order to study
the dependence of profile shape on frequency.
For example, there is a small ``bump" of emission beginning
at a phase of $\sim$ 350\de in Figure 3a, which we did not
address in this fitting process.  
After fitting the 1.4-GHz profile 
we fit the 430-MHz profile (shown in Fig. 3b)
by holding the positions and widths of each of these 1.4-GHz components
constant and varying only amplitudes.
Some features of the 430-MHz profile are left unmodeled under this
scheme. The largest discrepancy is at the edge of the main peak located
at $\sim$ 70$\DEG$.  
This discrepancy could be eliminated by shifting
the position of component 1 earlier in phase.
This apparent spreading out of the components as we go to lower frequency
suggests that some of this emission may be ``conal".

MSPs are thought to have possibly complex magnetic field topologies
from their history of accretion \citep{Ruderman91}.  Furthermore
the work of \citet{Xilouris98} suggests that pulsars with compact
light cylinders are more likely to exhibit abnormal profile evolution
with respect to the Rankin (1983) model\nocite{Rankin83}.  PSR J0030+0451
does indeed have a relatively small light cylinder radius 
($\sim $ 230 km) and a complex field topology even among MSPs 
as measured by the total
number of Gaussian components required to model its profile.  PSR
J0030+0451 requires 6 components compared to 
the average
for MSPs, $4.2 \pm 0.4$, measured by Kramer et al. (1998)\nocite{Kramer98},
hereafter K98. 

We also calculate the 10\% and 50\% widths of the main pulse
for comparison with the table of pulse widths for MSPs from K98.
Note that we let all 18 parameters vary
at both 430 MHz and 1.4 GHz in order to do this calculation.
The 10\% and 50\% widths at 1.4 GHz
are $64\DEG \pm 1\DEG$ and $48\DEG\pm 1\DEG$ respectively.  These numbers fall in the middle
of the distribution for MSPs shown in K98.
At 430 MHz the 10\% and 50\% widths 
are $74\DEG \pm 5\DEG$ and $54\DEG \pm 5\DEG$ respectively. 

\subsection{Polarimetry and Geometry of MSP J0030+0451}\label{sec:0030polar}

The large duty cycle of PSR J0030+0451
with both main pulse and interpulse emission is a common characteristic of MSPs.
The availability of polarization
information over a wide range of phases (see Fig.~4) in theory
allows an accurate determination of the magnetic field inclination 
angle $\alpha$.  
Of particular interest is the measurable polarization
position angle (PPA)
at the trailing edge of the interpulse (from $\sim$4.0 ms to
about $\sim$4.4 ms in 
phase in Fig. 4b) 
in a region where emission is at its minimum.
(Note that a constant offset, equal to the level of emission from
4.5 to 4.9 and 0.0 to 0.2 ms of phase,
has been subtracted from these data in each of the 4 
Stokes parameters)
Either the emission from this region is nearly 100\% linearly polarized
or in subtracting a constant offset from our data we have subtracted
off what is significant
baseline emission from this pulsar.  

Using polarimetric information, it is possible in some cases to determine
whether observed emission is `core' or `cone' emission, thereby
giving additional clues about the geometry.
\citet{Rankin83} (also see Rankin 1993\nocite{Rankin93}) identifies
several properties of core emission that are present in component 3:
(1) the steepest (either positive or negative) slope in polarization
position angle is at the phase of components 3 and 4;
(2)  component 3 is significantly
linearly polarized; and (3) the circular polarization, although very weak, 
changes sign underneath
component 3.  
In the following paragraphs we look to RVM fits to the observed 
PPA profile for further information.

Two possible RVMs are shown in Figure 4, the first with \nAlph=8\de
and the second with \nAlph=62\nde.  The first thing to note is that
neither case is particularly appealing.  Moreover there is no RVM
that would fit the jaggedness of the PPA sweep near the main pulse
peak.
We consider the nearly aligned case first, shown with parameters
\nAlph=8\de and \nbet=1\de.   Actually fitting to the data returns
a range of parameters which depend on initial values, so this is not a fit
to the data but rather an example of a plausible model.
From this model we expect the emission to be primarily conal and
thus the center of symmetry actually should be at a flux
{\em minimum}.
We therefore place the center
of symmetry of the model, which represents the magnetic pole,
in between the pulse and interpulse.
The pulse and the interpulse represent two 
crossings of
a single very wide ``cone" of emission, and therefore we 
expect that all the components
will display properties of conal emission, 
which they do not (see preceding paragraph).

In the second case we attempted to make \nAlph\ as large as possible
while still providing a reasonable fit to the data.
The model shown has \nAlph=62\de and \nbet=10\de
(see Fig. 4). This model places the center of symmetry of the RVM 
at the trailing edge of the main pulse near to, but
not coincident with,
peaks 3 and 4.  This is consistent with the RVM's prediction
of core emission from this location.
Specifically we expect the center of symmetry of the
PPA is at the phase of the {\em core} emission with any {\em cone} emission
falling on either side of it in phase.  We in fact only see emission on
the leading edge of this core emission (i.e. we see no emission to the right
of component 4, only to the left).  Perhaps we only see one-half of
the ``cone", a situation discussed earlier in the case of PSR J1313+0931 
(see \S\ref{sec:1313}).  
Alternatively \citet{Hibschman99} show that corrections to the 
simple RVM, resulting from
the effects of a Goldreich-Julian current on the magnetic field and
aberration, will
produce various asymmetries in the polarization sweep that we could be 
seeing.

We conclude that the polarimetry data suggest a large \nAlph\ and
that we are likely seeing two poles.

\subsection {MSP J0030+0451 and the Local ISM}

This pulsar allows new measurements of the physical state of the
ionized component of the local ISM.
PSR J0030+0451 lies roughly in the middle of the quadrant of
the Galaxy from $l=90\DEG$ to $l=180\DEG$
which is devoid of any pulsars with independent distance estimates.
Ultimately we will able to determine
the mean density using the trigonometric parallax derived from
ongoing timing measurements. The result can be compared with
that from other pulsars 
within 1 kpc of the sun
(Toscano et al. 1999 \nocite{Toscano99} and references therein)
to improve the 3D map of the local ionized gas.

The diffractive scintillation parameters for J0030+0451 provide an estimate 
of the turbulent state of the local ISM on scales of $10^{10}$ cm
which can be compared to results of recent studies of nearby pulsars
\citep{Bhat98, Phillips92, Rickett00}.
We expect that J0030+0451 will exhibit {\em weak} scattering at 1.4 GHz
given its low DM of 4.3328 $\pm$ 0.0002 cm$^{-3}$pc. 
\citet{Backer75} estimates the critical
frequency at which the scintillation modulation index drops
below unity to be 2.4 GHz for this DM, although there are significant 
variations 
along different paths through comparable column densities.
Our observations indicate that J0030+0451 is indeed weakly scattered
at 1.4 GHz.  In order to measure the scintillation modulation index
we inspect the variation from epoch to epoch
of the signal to noise ratio (S/N) of our
1.4-GHz profiles. This is
a reliable measure of the modulation index owing to
the stability of the 1.4-GHz receiver and observing conditions.
Our dominant concern in using S/N as a measure of flux
variability is the possible correlation of S/N with
zenith angle.  However, a plot of S/N vs zenith angle for all of 
the 3-minute average profiles shows no correlation, and indeed
the documents characterizing the latest Arecibo upgrade 
show that the system temperature is expected to be nearly
constant for zenith angle $\lesssim 19\deg$  
\citep{arecibo86} 
and the gain is expected to decrease
by 2\% going from 15\de to 20\de in zenith angle
\footnote{also http://www.naic.edu/techinfo/teltech/statisti.htm}.
We see the signal to noise ratio (S/N)
vary from 4 to 13, with a mean of 10 and a standard deviation of 2.5,
yielding an estimate of 0.25 for the scintillation modulation index.
Flux measurements at both the
Arecibo Observatory and the VLA are consistent with this value.
At the VLA we obtained
two one-hour observations
on 1998 February 23 and 26 at 1.4 GHz.  
Flux transfer from  primary and secondary calibrators yielded
fluxes of $1.0 \pm 0.1$ mJy and $0.7 \pm 0.1$ mJy.
The typical flux we obtain from the Arecibo observations 
is $0.6 \pm 0.1 $ mJy at 1.4 GHz.
The range of these three values is consistent with
our measured scintillation modulation index of 0.25.
This value of the modulation index leads to an estimate of the 
amplitude of the electron-density power spectrum, $C_n^2 \sim 1.5 \times
10^{-4}$m$^{-20/3}$, using relation C2 in \citet{Rickett00}.

At 430-MHz we looked for scintillation effects across the 10-MHz band
which is divided into 32 channels.
In some scans we see a variation by
as much as a factor of 3 across the band, but always monotonically
changing across the bandpass. We conclude that the scintillation bandwidth 
is greater than, but on the order of 10 MHz.
We use this lower bound on the scintillation bandwidth to estimate
an upper bound on 
$C_n^2 \lesssim 7\times 10^{-5}$
m$^{-20/3}$ using relation C3 in \cite{Rickett00}.
$C_n^2$ depends on the pulsar's distance to the
$11/6$ power which leads to an uncertainty in $C_n^2$ of
a factor of 4 in either direction given variations in the local
electron density reported by \citet{Toscano99}.
We conclude that $C_n^2$ along the path the PSR J0030+0451
is probably in the range between the low values
reported by 
\citet{Rickett00} and \citet{Phillips92} toward B0809+74 and B09050+08
and the higher values reported by
\citet{Bhat98} toward a dozen different objects.

Temporal variations of the DM inform us about perturbations on larger scales
of $10^{14}$ cm.  Our multifrequency timing observations (4 days
of dual frequency observations) allow us to
place an upper limit on the DM gradient of 0.0003 cm$^{-3}$pc yr$^{-1}$.
This small value is consistent with the relationship of DM gradient to 
DM suggested by \cite{Backer93}.

\section {Discussion}\label{sec:disc}

We have confirmed three new pulsars in a search of the Arecibo
sky covering 470 square degrees with a sensitivity of
1 mJy.  The objects are solitary pulsars.

Our polarimetric study
of PSR J0030+0451 suggests an orthogonal or nearly orthogonal
model with a large \nAlph\ .  
Multi-frequency (as opposed to dual-frequency) 
observations of PSR J0030+0451 are needed to fully assess the
component evolution and thereby determine whether it
is an orthogonal or an aligned rotator.  

PSR J0030+0451 provides a useful new probe of the plasma in the
local interstellar medium.

Of the approximately dozen MSPs which display interpulses
PSRs B1012+5307, B1855+09, and J2322+2057 
\citep{Nicastro95, Segelstein86, Nice93, Xilouris98} 
bear striking
resemblance to PSR J0030+0451: Pulse periods are between
4.8  ms and 5.4  ms and the magnetic fields are between 
2.2 $\times 10^8$ G and 3.2 $\times 10^8$ G. 
Each one is thought to be an orthogonal rotator, and in each case
the interpulse {\em increases} in amplitude compared to the main
pulse at higher frequency. \citet{Xilouris98} suggests
that the interesting properties of this group of pulsars
are due to a quadrupolar moment of the magnetic field becoming
important, and that outer-gap emission also is visible.

Of the 56 known MSPs in the disk of the Galaxy \citep{Camilo99,Edwards00,
Lyne00, Manchester00} 
PSR J0030+0451 is the 9th solitary MSP (i.e., it is not
in a binary system).  
These objects present particular requirements on the standard evolutionary
scenario of MSPs.
If the neutron star is spun up via accretion
by mass transfer from a companion star, the companion must somehow be 
obliterated.  Various solutions
have been proposed such as a common envelope evolution whereby during
the accretion phase, the core of the companion and the neutron star 
spiral in and
finally collapse into a single star \citep{vandenheuvel95}.  
Alternatively the new MSP may evaporate or ablate its companion as
is suggested in observations of PSR B1957+20 \citep{Fruchter90, Zaven94},
and PSR B1744$-$24A \citep{Lyne90}.  Neither of these solutions
is manifest
in observations, and the evolutionary history of
solitary MSPs remains an unsolved problem.   

\acknowledgements
We wish to thank Michael Kramer
for the use of his Gaussian fitting program,
and Christophe Lange for his TOA uncertainty code.
We also thank
Shauna Sallmen for her expertise in polarimetry measurements and for use
of her software.
We are grateful to the Arecibo telescope operators for running the
search observations and for their 
assistance in the follow-up timing observations.
We thank Alex Wolszczan for use of the Penn State Pulsar Processor, 
Fernando Camilo for comments on an earlier version of this manuscript, and
Duncan Lorimer and Dan Stinebring for assistance in 
observations.   Lastly we thank Dick Manchester, the referee, whose
comments significantly improved the manuscript.

\begin{table}
\caption[]{Declination coverage of the Berkeley/Cornell
Drift Search}
\begin{center}
\begin{tabular}{lrrr}
\tableline
DEC &   Area Observed  & 2+ Observed \cr
    &   (sq.deg.) & (sq.deg.) \cr
\tableline
-0.5\de   &  42 (12\%)   &   10 \cr
 4.5\de   &  103 (29\%)   &  36  \cr
 9.5\de   &  146 (41\%)   &  95  \cr
24.5\de   &  49 (15\%)   &   15  \cr
29.5\de   &  79 (25\%)   &  25  \cr
34.5\de   &  51 (17\%)   &   10  \cr
\tableline
\end{tabular}
\end{center}
\end{table}

\begin{table}
\caption[Pulsar Parameters] {Observed Pulsar Parameters\tablenotemark{a}.}
\begin{tabular}{||l|c|c|c||} \tableline
     &				J0030+0451   &   J0711+0931   & J1313+0931 \\ \tableline
Right Ascension (J2000) &		00\hou30\minu27\seco432(4)	  &  07\hou11\minu36\seco18(2) & 13\hou13\minu23\seco0(1) \\ \tableline					
Declination (J2000) &		04\de 51\pr 39.7\dpr(1)  &  09\de 31\pr 25\dpr(1)  & 09\de 31\pr 56\dpr (1) \\ \tableline
Galactic longitude & 113.1\de & 206.7\de  & 320.4\de \\ \tableline
Galactic latitude & $-$57.6\de    &  8.7\de & 71.7\de \\ \tableline
Period(s) & 	0.004865453207369(1) & 1.21409049248(1) & 0.84893275073(2) \\ \tableline
Period derivative $(10^{-15}$ s s$^{-1})$&   1.0(2)$\times 10^{-5} $ &  0.4(1)  &  0.8(1) \\ \tableline
Epoch (MJD) &				50984.4 &  51199.5  &         50984.5 \\ \tableline
Dispersion Measure (pc cm$^{-3}$) &            4.3328(2) & 45.0(1)      &   12.0(1) \\ \tableline
Timing data span (MJD) &		50789--51622 & 50789--51622  & 50788--51080 \\ \tableline
Number of epochs of data\tablenotemark{b} & 9 & 7 & 5 \\ \tableline
Number of days of data & 28 & 21 & 18 \\ \tableline
RMS timing residual \tablenotemark{c} & 5 $\mu$s   & 6 ms      &  1 ms \\ \tableline
Flux density $S_{430}$ (mJy) &		7.9(2)		& 2.4(1)  &       3.5(1)  \\ \tableline
Flux density $S_{1400}$ (mJy) &		0.6(2)    &  0.04(1) &  0.16(1)  \\ \tableline
DM distance (pc)\tablenotemark{d} &		230 & 2450  & 780 \\ \tableline
$L_{400}($mJy kpc$^{2}$) & 0.4  & 14 & 2.1\\ \tableline
Spectral index & $-2.2 \pm 0.2$ & $- 3.5 \pm 0.3$ & $ - 2.6 \pm 0.2$ \\ \tableline
Characteristic age (y) &  $ 7.8 \times 10^9 $ &  $ 4.9 \times 10^7 $  & $1.7\times 10^7 $ \\ \tableline
Magnetic field strength (G)\tablenotemark{e} & $ 2.2 \times 10^8 $ & $ 7.1 \times 10^{11} $  & $ 8.3 \times 10^{11} $\\ \tableline 
\end{tabular} 
\tablenotetext{a}{Uncertainties in parenthesis refer to the last digit quoted.} 
\tablenotetext{b}{We define `epoch' as data separated by 2 weeks or more}
\tablenotetext{c}{RMS of residuals using 3 minute averages}
\tablenotetext{d}{Model from \citet{TaylorCordes93}} 
\tablenotetext{e}{$ B_o = 3.2 \times 10^{19} G \sqrt{P({\rm s})\dot{P}} $}
\end{table}

\begin{table}
\caption[Table 3]{Gaussian Component Parameters of PSR J0030+0451.}
\begin{center}
\begin{tabular}{|l|r|r|r|r|r|r|r|}
\tableline
Peak \# & 1 &    2 &     3 &     4 &     5 &     6 \\
\tableline
Center (deg) &78.7 & 88.1 & 95.9 & 107.9 &  253.1 & 271.5 \\
Width (deg) & 16.0 & 3.9 & 5.2 & 7.0 &   32.8 & 12.2 \\
Amplitude (430 MHz) & 1.95 & 1.00 & 2.35 & 2.62 &  0.363 &  0.162  \\    
Amplitude (1.4 GHz) & 1.15  & 1.00 & 1.25 & 1.67 & 0.311 &  0.444 \\
\tableline
\end{tabular}
\end{center}
\end{table}

\clearpage
\setcounter{figure}{0}

\begin{figure}
\plotone{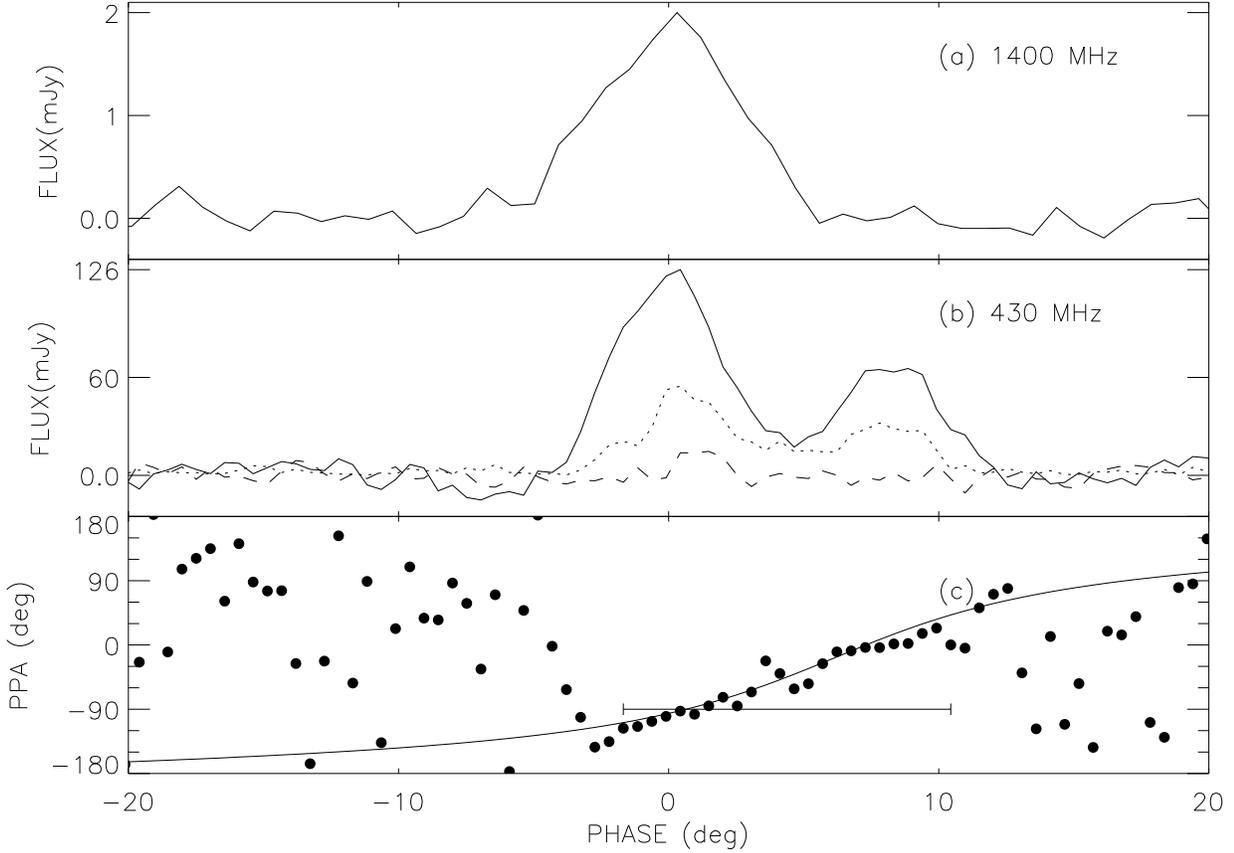}
\caption{PSR J0711+0931 pulse profiles:  
(a) 1.4 GHz intensity vs. phase;  (b) 430 MHz intensity vs. phase; and  
(c) polarization position angle (PPA) at 430 MHz.  In (b) the solid 
line shows total intensity, the dotted line shows
linear polarization, and the dashed line shows circular polarization.  The
sense of circular polarization is $V = S_{left} - S_{right}$, 
the IEEE convention.  The zero of PPA is arbitrary.  In
(c) the solid line shows an
RVM superimposed on the PPA data.
The model shown was fit to the data with $ \alpha $ set to
$ 60.0 \DEG $. The best fit $ \beta $ in this case was $-7.2 \DEG \pm 2\DEG$.  The
horizontal bar shows the fitting region used.  A
number of values of \BAlph and \bet work equally well (see text for details).}
\end{figure}

\begin{figure}
\centering{\includegraphics{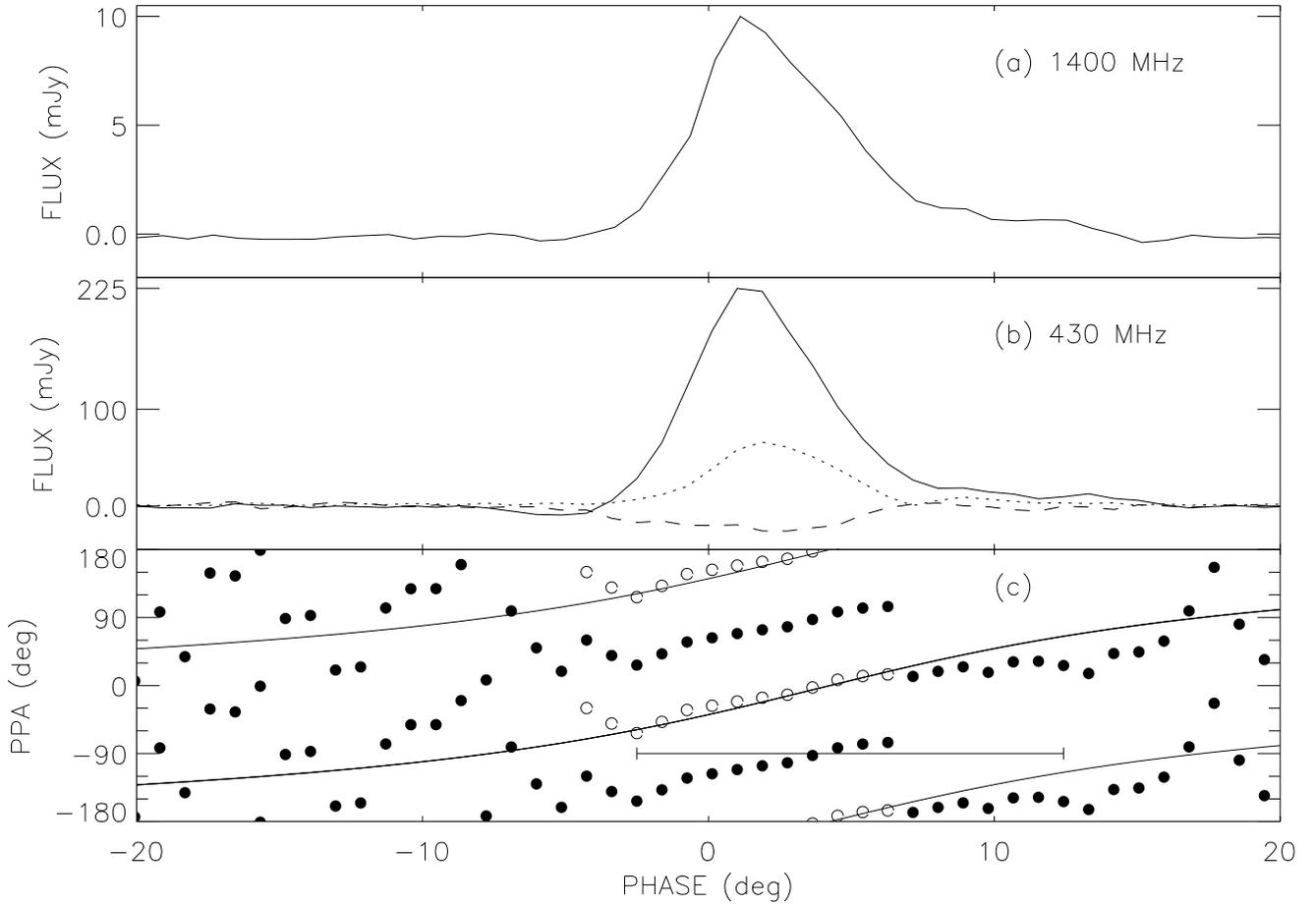}}
\caption{PSR J1313+0931 pulse profiles:  (a) 1.4 GHz intensity
vs. phase;  (b) 430\Mhz intensity vs phase; and (c) polarization position
angle (PPA) at 430 MHz.  In (b)  
the solid line shows total intensity, the dotted line shows
linear polarization, and the dashed line shows circular polarization.  In
(c) the open circles show the data after being rotated to account for
the orthogonal mode.  The rotation was implemented up until 7\de
phase, where we suggest the orthogonal transition happens.
In (c) the solid shows a RVM superimposed on the PPA data.
The model shown below was fit to the data with $ \alpha $ set to
$ 60.0 \DEG $. The best fit $ \beta $ in this case was $-9.5 \DEG \pm 3\DEG$.  The
horizontal bar shows the fitting region used.  A
number of values of \BAlph and \bet work equally well (see text for details).}
\end{figure}

\begin{figure}
\plotone{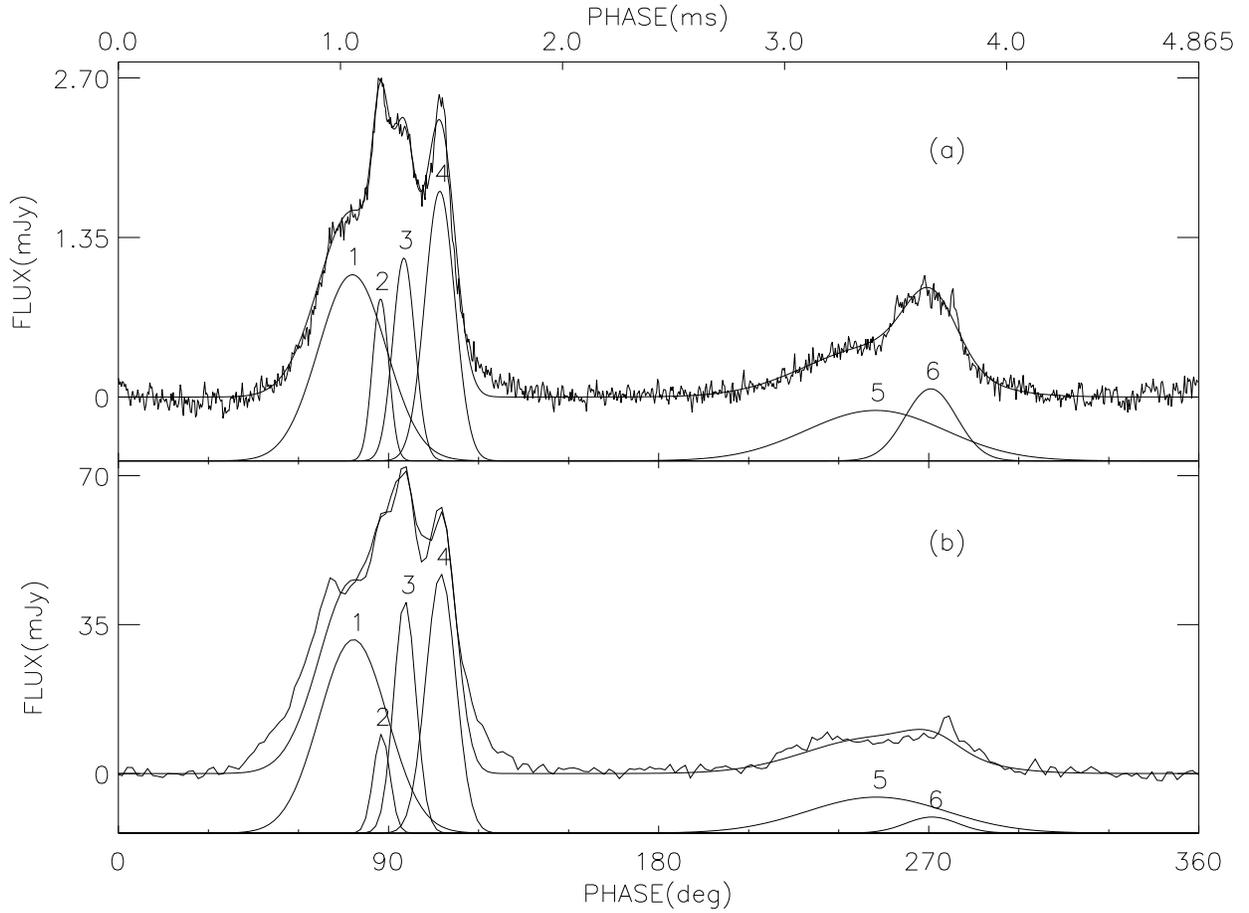}
\caption{Decomposition of PSR J0030+0454 intensity profile into 6 components:  
(a) 1.4 GHz and (b) 430\Mhz.  The best-fit Gaussian components at 1.4 GHz
were only allowed to vary in amplitude in order to fit the 430\Mhz profile.
See text and Table 3 for details.}
\end{figure}
 
\begin{figure}\label{fig:0030polar}
\plotone{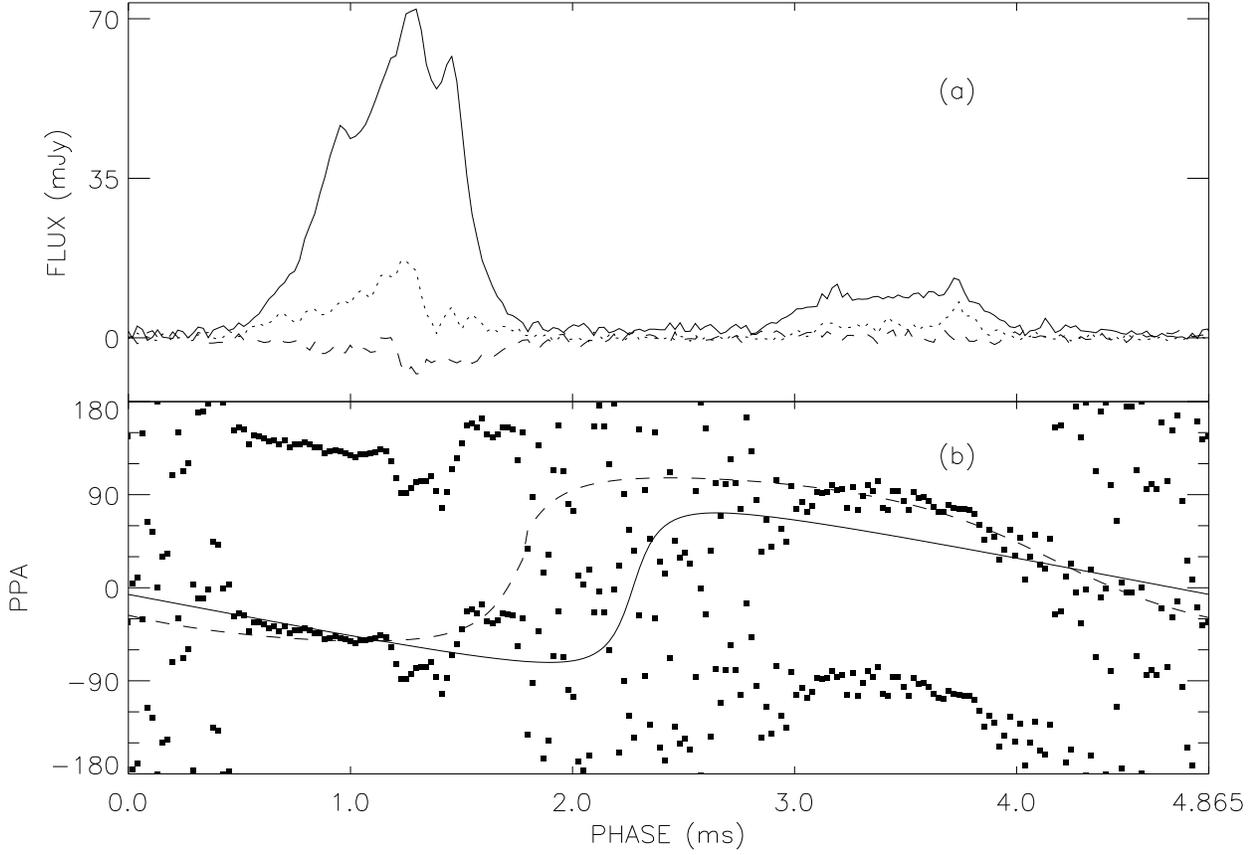}
\caption{PSR J0030+0454 at 433 \Mhz.  In panel (a) 
the solid line shows intensity vs. phase, the dotted line shows
linear polarization, and the dashed line shows circular polarization.   
In panel (b) two RVM's are superimposed on the PPA data.
The solid line demonstrates the best-fit scenario in the case of
the nearly aligned rotator (see text) which is
drawn for 
$ \alpha=8 \DEG $ and $ \beta =1 \DEG $.  
The dashed line demonstrates the best-fit scenario in the case of
a more orthogonal rotator and has values
$ \alpha=62 \DEG $ and $ \beta =10 \DEG $. }  
\end{figure}

\end{document}